# $^{12}$B($n,\gamma$)$^{13}$B REACTION AS ALTERNATIVE PATH TO ASTROPHYSICAL SYNTHESIS OF $^{13}$C ISOTOPE

**Dubovichenko S.B.**[1,2,*], **Burkova N.A.**[2], **Dzhazairov-Kakhramanov A.V.**[1,*], **Yertaiuly A.**[2]

[1]*Fesenkov Astrophysical Institute "NCSRT" ASA MDASI Republic of Kazakhstan (RK), Observatory 23, Kamenskoe plato, 050020, Almaty, RK*
[2]*Al-Farabi Kazakh National University, 050040, Almaty, RK*

The total cross sections of the neutron radiative capture on $^{12}$B at astrophysical energies to the ground state of $^{13}$B have been calculated in the energy range of $10^{-8}$ to 10 MeV within the framework of a modified potential cluster model with the classification of orbital states according to Young diagrams. Reaction rates in the temperature range of 0.01 to 10 $T_9$ and their analytical parameterization were obtained. The calculated rates of $^{12}$B($n,\gamma$)$^{13}$B excess the previous results by approximately to one order. Cross sections and reaction rates of $^{12}$C($n,\gamma_{0+1+2+3}$)$^{13}$C are calculated and compared to the $n^{10}$B, $n^{11}$B, $n^{12}$B, and $p^{12}$C reaction rates. It is proposed that obtained rates of the $^{12}$B($n,\gamma$)$^{13}$B reaction should be taken into account in *novel* scenarios of stable isotope $^{13}$C synthesis without of $^{12}$C hydrogen burning.

## 1. Introduction

The study of the synthesis of *stable elements* in astrophysical processes, start with the Big Bang and the further evolution of the Universe is undoubtedly acknowledged both with the fundamental and practical interests. Namely, it is important to clarify what their accumulative balance in planetary systems can be recognized as possible sources of energy and fossil minerals in the distant future. The reaction chains that lead to the formation of the stable elements include the processes involving *short-lived radioactive isotope,* which is one of the most actual item of our days.

So that, the reviews [1,2] discuss the scenarios for the synthesis of neutron-rich light nuclei in supernova explosions as a key door of the BBN mechanisms. In particular, in [1] a chain involving boron isotopes includes $^8$B, $^{10-15}$B, $^{17}$B and $^{19}$B. At present, isotopes $^{20,21}$B have been synthesized [3] and the most complete information is presented in the *ENSDF* database [4]. To implement the calculation scheme for evaluating the *r*-processes proposed in [1], it is necessary to have reaction rates of the processes

$$^{10}\text{B}(n,\gamma)^{11}\text{B}(n,\gamma)^{12}\text{B}(n,\gamma)^{13}\text{B}(n,\gamma)^{14}\text{B}(n,\gamma)^{15}\text{B}(e,\nu)^{15}\text{C} \ etc.$$

At present, there are experimental data for the total cross sections for the neutron radiative capture $^{10,11}$B($n,\gamma$) [5–7], as well as preliminary data for the $^{13,14}$B($n,\gamma$) reaction [8].

Today, macroscopic calculations of stellar dynamics, where the process $^{12}$B($n,\gamma$)$^{13}$B is included also, are based only on the calculations of the reaction rate of 25 years old work [9], and later one [10]. In these works, the total cross section $\sigma = \sigma_{DC} + \sigma_{res}$ is represented by the sum of the direct capture (DC) cross section of the $\sigma_{DC}$ and the resonance process $\sigma_{res}$. The latter is calculated by the Breit-Wigner method, that is, it depends on the accuracy of the corresponding $\Gamma$-widths. Resonance cross sections $\sigma_{DC}$ in

---

[*] Corresponding authors: albert-j@yandex.kz, dubovichenko@gmail.com

[9] are calculated in a model-less way. In [10], the $\sigma_{DC}$ is calculated in a cluster model, and the interaction potentials are fitted using the experimental cross sections of the $^{12}$B($d,p$)$^{13}$B process.

Here, we study the reaction of the neutron radiative capture $^{12}$B($n,\gamma$)$^{13}$B. The peculiarity of this process is in the fact that the isotopes of $^{12}$B($\beta^-$)$^{12}$C and $^{13}$B($\beta^-$)$^{13}$C, i.e. the participants in the initial and final channels, are short-living ones with a lifetime of 20.20 ms and 17.33 ms respectively. Their disintegration with a 100% probability leads to the formation of stable carbon nuclei. However, the ignition of these processes should be provided by the processing of $^{12,13}$B isotopes, for example within the above chain. As we aim on the $^{13}$B production, the preliminary remark is pertinent: due to the short lifetime of $^{12}$B, the direct measurements of cross sections, for example, $^{12}$B($n,\gamma$)$^{13}$B reaction are very difficult. Therefore, the theoretical calculations of such reactions become relevant.

Besides, studying the $^{12}$B($n,\gamma$)$^{13}$B capture has the various attitude towards examination of the ratio $^{12}$C/$^{13}$C in the different stellar conditions [11–14]. In particular, it concerns stars at different stages of formation. The ratio $^{12}$C/$^{13}$C values for different masses of stellar objects is changed of 20.5±1.5 for object α Ari by mass (M☉) equals 1.0 to 6±1 for object α Ori by mass (M☉) equals 15.0 [13]. Also according to [11] the elemental ratio $^{12}$C/$^{13}$C is equal to 60–70 and higher. The solar ratio is equal to 90 [11], according to [14] the observed solar ratio by mass is (C/$^{13}$C)☉ = 82.2.

In our approach, the modified potential cluster model (MPCM) is used for the $^{12}$B($n,\gamma$)$^{13}$B reaction, within the framework of which reliable results of calculating the total cross sections and reaction rates for $^{10,11}$B($n,\gamma$) capture were obtained (see, for example, [15]). In general, the possibility of describing the main characteristics of bound states (BS) of certain nuclei, and astrophysical $S$-factors or the total radiative capture cross sections of more than 30 reactions at thermal and astrophysical energies is shown in works [16,17]. The calculations of these processes were carried out on the basis of the modified version of the potential cluster model (PCM) with the classification of orbital states according to Young diagrams and forbidden states (FS).

The certain success of the MPCM can be explained by the fact that the partial potentials of intercluster interaction are constructed not only on the basis of the known elastic scattering phase shifts, but also taking into account the classification of cluster states according to Young diagrams [18,19]. The potentials of the bound states with the FS in certain partial waves are constructed basing on the description of the binding energy, mean square radii, and asymptotic constants (AC) of the nucleus in the certain cluster channel. Furthermore, these potentials make it possible to calculate some characteristics of the interaction of the particles in the processes of elastic scattering and reactions. For example, these may be astrophysical $S$-factors of radiative capture reactions or total cross sections of such reactions.

**2. Calculation methods**

We present here some definitions and formalism elements that necessary for further discussions. The total cross sections of the radiative capture $\sigma(NJ,J_{\mathrm{f}})$ for $EJ$ and $MJ$ transitions in the PCM are given, for example, in [20] or [16,17] and have the form

$$\sigma_c(NJ, J_f) = \frac{8\pi K e^2}{\hbar^2 q^3} \frac{\mu(J+1)}{(2S_1+1)(2S_2+1)} \frac{A_J^2(NJ,K)}{J[(2J+1)!!]^2} \sum_{L_i,J_i} P_J^2(NJ, J_f, J_i) I_J^2(J_f, J_i), \quad (1)$$

where σ is the total cross section of the radiative capture process, μ is the reduced mass of particles of the initial channel, $q$ is the wave number of particles of the initial channel, $S_1$ and $S_2$ are the spins of particles in the initial channel, $K$, $J$ is the wave number and multipole momentum of the γ-quantum in the final channel, $NJ – E$ or $M$ transitions of the $J^{th}$ multipolarity from the initial $J_i$ state of two particles in the continuous spectrum to the final bound $J_f$ state of the nucleus in the two-cluster channel.

For the electric orbital $EJ(L)$ transitions, values $P_J$, $A_J$ and $I_J$ defined as (see [16,17])

$$P_J^2(EJ, J_f, J_i) = \delta_{S_i S_f} \left[(2J+1)(2L_i+1)(2J_i+1)(2J_f+1)\right] \left(L_i 0 J 0 | L_f 0\right)^2 \begin{Bmatrix} L_i & S & J_i \\ J_f & J & L_f \end{Bmatrix}^2, \quad (2)$$

$$A_J(EJ, K) = K^J \mu^J \left[\frac{Z_1}{m_1^J} + (-1)^J \frac{Z_2}{m_2^J}\right], \qquad I_J(J_f, J_i) = \langle \chi_f | r^J | \chi_i \rangle.$$

Here, $L_f$, $L_i$, $J_f$, $J_i$ are the orbital and total momenta of the particles of the initial (*i*) and final (*f*) channel, $m_1$, $m_2$, $Z_1$, $Z_2$ are the masses and charges of the initial channel particles, $I_J$ is the integral of the initial $\chi_i$ and final $\chi_f$ wave functions states as a function of the relative motion of clusters with intercluster distance $r$.

To consider the magnetic $M1(S)$ transition due to the spin part of the magnetic operator, we use the relations ($L = L_i = L_f$)

$$P_1^2(M1, J_f, J_i) = \delta_{S_i S_f} \delta_{L_i L_f} \left[S(S+1)(2S+1)(2J_i+1)(2J_f+1)\right] \begin{Bmatrix} S & L & J_i \\ J_f & 1 & S \end{Bmatrix}^2, \quad (3)$$

$$A_1(M1, K) = \frac{\hbar K}{m_0 c} \sqrt{3} \left[\mu_1 \frac{m_2}{m} - \mu_2 \frac{m_1}{m}\right], \qquad I_J(J_f, J_i) = \langle \chi_f | r^{J-1} | \chi_i \rangle, \quad J = 1,$$

where m is the mass of nucleus, $\mu_1$ and $\mu_2$ are magnetic momenta of the clusters, which taken from [21,22], namely $\mu_n$ = -1.9130427$\mu_0$ and $\mu(^{12}B)$ = 1.00272$\mu_0$.

In general for the $MJ$ transitions this expression is written as

$$P_J^2(MJ, J_f, J_i) = \delta_{S_i S_f} \left[S(S+1)(2S+1)(2J_i+1)(2L_i+1)(2M+1)(2J+1)(2J_f+1)\right] \cdot$$
$$\cdot (L_i 0 L 0 | L_f 0)^2 \begin{Bmatrix} L_i & M & L_f \\ S & 1 & S \\ J_i & J & J_f \end{Bmatrix}^2, \quad (4)$$

where $M = J – 1$, $I_J(J_f, J_i)$ defined in (3) for arbitrary $J$ and

$$A_{\text{J}}(MJ,K) = \frac{\hbar K}{m_0 c} K^{J-1} \sqrt{J(2J+1)} \left[ \mu_1 \left(\frac{m_2}{m}\right)^J + (-1)^J \mu_2 \left(\frac{m_1}{m}\right)^J \right]. \quad (5)$$

In carried out calculations the following values of particle masses was used $m_n$ = 1.00866491597 amu [21], $m(^{12}B)$ = 12.014352 amu [23], and $\hbar^2/m_0$ constant is equal to 41.4686 MeV·fm$^2$, where $m_0$ is the atomic mass unit (amu). Methods for calculating other quantities within the MPCM, for example, mean square radii or binding energies, which are used further, are described in [16,17].

In the present calculations of the neutron radiative capture on $^{12}$B the $E$1 transitions was taken into account. Such transitions in $n^{12}B \to {}^{13}B\gamma$ process are possible between doublet $^2S_{1/2}$ or quartet $^4S_{3/2}$ states of scattering and the ground $^{2+4}P_{3/2}$ bound state of $^{13}$B nucleus in the $n^{12}$B channel, which can be mixed by 1/2 and 3/2 channel spins.

### 3. Physical backgrounds of construction the potentials in the MPCM

Let us consider in more detail the procedure for constructing intercluster partial potentials (in this case, the interaction potentials of the neutron and the target nucleus) for a given orbital momentum $L$ and other quantum numbers. Definitions of the criteria and the sequence for finding the corresponding parameters are described in [24].

First of all, we find the parameters of the bound state potentials, in particular of the $n^{12}$B GS, which for a certain number of the bound AS and FS in the given partial wave are fixed unambiguously by the binding energy and AC in the considered channel. While constructing the partial interaction potentials in the MPCM it is assumed, that they depend not only on the orbital momentum $L$, but also on the total spin $S$ and the total momentum $J$. Here and earlier [16,17] we use the Gaussian form of the interaction potentials, which depends on the quantum numbers $JLS$, and in some cases on the Young diagrams {$f$} [19].

In other words, for different momenta of $JLS$ we will have different values of the parameters for the partial potentials. Since $E$1 or $M$1 transitions between different states of $^{(2S+1)}L_J$ in the continuous and discrete spectra are usually considered, the potentials of these states will be different [16,17]. Note that we use the MPCM for our calculations, and one of its modifications consists in the explicit dependence of the potentials used on the Young diagrams {$f$} and, in some cases, the mixing of scattering states according to Young diagrams may appear [16,17,19]. So that the explicit dependence of the interaction potentials on the Young diagrams makes it possible to use different potentials in the scattering states and discrete spectrum if they depend on a different set of such diagrams [16,17,24].

The accuracy, of the parameters of the BS potential is determined by the accuracy of the AC. The potential does not contain other ambiguities, since the classification of states according to Young diagrams makes it possible the unique fixing of the number of the bound forbidden and allowed states (AS) in a given partial wave. Their number completely determines the depth of the potential, and its width depends entirely on the value of AC. The principles for determining the number of FS and AS in the given partial wave based on Young diagrams for the $n^{12}$B system are given below.

It should be noted, that the results of the calculations of the charge radius depend on the model wave functions (WF), which leads to some model error. At the same time, AC values are determined by the asymptotics of radial WFs at one point and, apparently, contain a significantly smaller error. Therefore, the GS and BS potentials are constructed so that they are consistent with the AC values obtained on the basis of the independent methods that allow one to extract AC from the experimental data (see, for example, [25]).

The intercluster potential of the nonresonance scattering process by the scattering phase shifts for a given number of BSs allowed and forbidden in the partial wave is also constructed quite unambiguously. The accuracy of determining the parameters of such a potential is primarily associated with the accuracy of extracting the scattering phase shifts from the experimental data and can reach 20–30%. Here such potential does not contain ambiguities, since the classification of states according to Young diagrams makes it possible to unambiguously fix the BS number, which completely determines its depth, and the potential width at a given depth is determined by the shape of the scattering phase shift.

When constructing a nonresonance scattering potential from the data on the spectra of the nucleus, it is difficult to evaluate the accuracy of finding its parameters even for a given number of BSs. Such a potential, as is usually assumed for the energy range up to 1–3 MeV, should lead to the scattering phase shift close to zero or giveы a smoothly decreasing phase shift shape, since there are no resonance levels in the spectra of the nucleus.

In the analysis of resonance scattering, when a relatively narrow resonance with a width of the order of 10–100 keV is existed in the considered partial wave at the energies up to 1–3 MeV, the potential is constructed completely unambiguously. For a given number of BSs its depth is completely fixed by the resonance energy of the level, and its width is determined by the width of such a resonance. The error of its parameters usually does not exceed the error in determining the width of a given level. Moreover, this also applies to the construction of the partial potential from the resonance scattering phase shifts and the determination of its parameters from the resonance in the nuclear spectra.

### 4. Classification of cluster states

First, note that the state with $L = 1$ in the spin-mixed $^{2+4}P_{3/2}$ wave corresponds to the ground bound allowed state of $^{13}$B in the $n^{12}$B channel with the momentum and isospin of $J^\pi, T = 3/2^-, 3/2$ ($J^\pi, T = 1^+, 1$ is known for $^{12}$B [22]) at the binding energy of -4.878 MeV [26].

Furthermore, we consider a different, compared with the results of our previous article [27], classification of states according to Young diagrams. This leads to a different structure of the FS and AS and to other potentials. In addition, in the previous work, we considered that the GS of $^{13}$B in the $n^{12}$B channel is only the $^2P_{3/2}$ doublet level, but in this case we consider it the mixture of doublet and quartet waves, denoting it as $^{2+4}P_{3/2}$.

Here we choose for the GS of $^{12}$B the orbital diagram {4431} obtained in [27] for $^{12}$B when considering the $n^{11}$B channel. In this case, for $^{13}$B in the $n^{12}$B channel we have {1} + {4431} = {5431} + {4441} + {4432}. The first diagram is forbidden and has an orbital momentum $L = 1,2,3$. The second ($L = 1,3$) and the third ($L = 1,2$) diagrams are apparently allowed. Thus, the GS of $^{13}$B can be matched with both diagrams {4441} and {4432}. An unambiguous conclusion is not possible due to the lack of the production tables of Young diagrams for the number of particles of $N = 13$. Previously we used tables [28] for $N$ less than or equal to 8. It is possible that the scattering states are mixed according to these diagrams, as in the case for lighter systems [19]. Here, for certainty, we will use the diagram {4441} for the GS with a low-lying FS for the diagram {5431} and assume that the state for {4432} is located above and can be unbound. The scattering states also will be considered as dependent only from one diagram {4441}.

Let us turn to the description of the results for the potential of the GS of $^{13}$B in the n$^{12}$B channel is the $^{2+4}P_{3/2}$ wave with the FS. Such a potential should correctly reproduce the binding energy of $^{13}$B in the n$^{12}$B channel that is equal to -4.878 MeV, the AC, and it is acceptable to describe the rms charge and mass radii of $^{13}$B, for which the radii are 2.48(3) fm and 2.41(5) fm [29]. The charge and mass radii $^{12}$B are given in the same work and are 2.31(7) fm and 2.41(3) fm, respectively. The charge radius of the neutron is assumed to be zero, and its mass radius is taken to be equal to the proton radius of 0.877(5) fm [21].

The BS potentials should correctly describe the known AC values, which is associated with the asymptotic normalization coefficient (ANC) usually extracted from experiment as follows [25]

$$A_{NC}^2 = S_f \cdot C^2, \qquad (6)$$

where $S_f$ is the spectroscopic factor of the channel and $C$ is the dimensional AC, expressed in fm$^{-1/2}$ and defined from

$$\chi_L(r) = C \cdot W_{-\eta L+1/2}(2k_0 r), \qquad (7)$$

which is related to the dimensionless AC noted as $C_w$ [30], that we usually used. So that, $C = \sqrt{2k_0} C_w$, and dimensionless constant $C_w$ is defined by expression [30]

$$\chi_L(r) = \sqrt{2k_0} \cdot C_w \cdot W_{-\eta L+1/2}(2k_0 r). \qquad (8)$$

Here $\chi_L(r)$ is the numerical wave function of the bound state obtained from the solution of the radial Schrödinger equation and normalized to unity, $W_{-\eta L+1/2}$ is the Whittaker function of the bound state, which determines the asymptotic behavior of the WF and is the solution of the same equation without a nuclear potential, $k_0$ is the wave number corresponding to the channel binding energy, η is the Coulomb parameter equals zero in this case, and $L$ is the orbital momentum of the given bound state.

Theoretical results are known on the asymptotic normalization coefficients and spectroscopic factors $S_f$ for $^{13}$B in the n$^{12}$B channel from [31]. For the ANC at the channel spin of 1/2 the value 1.74 fm$^{-1}$ is given, and for the spectroscopic factor 0.56–0.60 at $\bar{S} = 0.58$. For the ANC at the channel spin of 3/2 the value 0.31 fm$^{-1}$ is given and for the spectroscopic factor 0.11–0.13 at $\bar{S} = 0.12$. For the total ANC values was obtained 2.05 fm$^{-1}$, for the spectroscopic factor 0.67–0.72 at $\bar{S} = 0.70(3)$, that we will use further, because the GS is considered as the mixture of these states.

Furthermore, on the basis of expression (6) for the GS, we find $\bar{A}_{NC}/\sqrt{\bar{S}} = \bar{C} =$

1.71 fm$^{-1/2}$, and since $\sqrt{2k_0} = 0.967$ then the dimensionless AC defined as $C_w = C/\sqrt{2k_0}$, and equals to $\overline{C}_w = 1.77$. However, the spectroscopic factor has an interval of 0.67–0.73; therefore, the value $C_w$ is laid within 1.75–1.81. In addition to the given above results [31], for the $^2P_{3/2}$ we can get $\overline{C}_w = 1.79$, and for the $^4P_{3/2}$ we find $\overline{C}_w = 1.66$.

If we use the value 1.1(3) given in [10] for the total spectroscopic factor, then for ANC 2.05 fm$^{-1}$ we get $\overline{C}_w = 1.41$, and the interval of dimensionless AC that takes into account the errors of the spectroscopic factor, is equal to 1.25–1.66. As a result, the interval of possible values of dimensionless constant is turned to be $\overline{C}_w = 1.25$–1.81 or $C_w = 1.53(28)$.

## 5. Construction of n$^{12}$B interaction potentials

Following [16,17], for the $n^{12}$B interaction potential in each partial wave with a given orbital momentum $L$, we use a potential of a Gaussian type with an explicit dependence on the $JLS$ momenta of the $n^{12}$B system

$$V(r, JLS) = -V_0(JLS)\exp(-\gamma_{JLS}r^2). \quad (9)$$

For the $P_{3/2}$ GS of $^{13}$B considered here in the $n^{12}$B channel, we find the parameters

$$V_{g.s.} = 245.016033 \text{ MeV}, \quad \gamma_{g.s.} = 0.275 \text{ fm}^{-2}. \quad (10)$$

The potential with the FS leads to the binding energy of -4.87800 MeV with an accuracy of the finite difference method (FDM) of 10$^{-5}$ MeV [16,17], the root-mean-square charge radius $R_{ch} = 2.42$ fm and the mass radius $R_m = 2.39$ fm. The AC value in the interval 6–16 fm turned out to be 1.81(1). The AC error shown here is determined by its averaging over the indicated distance interval.

For comparison, we give one more option of the GS potential, narrower than the previous one, corresponding to the AC located on the other edge of the given above range of possible values

$$V_{g.s.} = 385.26410 \text{ MeV}, \quad \gamma_{g.s.} = 0.45 \text{ fm}^{-2}. \quad (11)$$

It leads to the binding energy of -4.87800 MeV, charge radius of 2.42 fm, mass radius of 2.35 fm, and AC is 1.33(1) in the range of 5–16 fm.

Turning to the description of scattering processes, we note that as the potential of $S$ waves of the $n^{12}$B scattering, since they do not contain FS, we can take the zero depth parameter, which leads to zero scattering phase shifts. The GS potentials of the state have been used for the $P_{3/2}$ scattering wave, therefore the M1 transitions from $^{2+4}P_{3/2}$ scattering states to the $^{2+4}P_{3/2}$ GS turn out to be forbidden due to the orthogonality of wave functions, and in real numerical form they are 3–5 orders smaller than other transitions.

Here we assume that the resonance at 510 keV (excitation energy of 5.388(6) MeV [26]) is in the $D$ wave, two lower-lying resonances at 146 keV (excitation energy 5.024(6) MeV [26]) and 228 keV (excitation energy 5.106(10) MeV [26]) are present in the *P* waves. In this case, this is nothing more than an assumption, since there are no unambiguous quantum numbers for these states (see, for example, [26] or [32]). By the way, we could not construct the potential for resonance at 510 keV in the *P* wave – any parameters do not allow one to obtain such a small resonance width, but in the *D* wave this resonance is obtained automatically. All potentials of the *P* and *D* scattering waves have bound FS, GS potential has AS and FS and potentials of the *S* scattering waves have not FS.

Table 1. Characteristics of possible transitions from the initial $\{^{(2S+1)}L_J\}_i$ state to different components of the WF $\{^{(2S+1)}L_J\}_f$ of the GS of $^{13}$B at the neutron capture on $^{12}$B.

| No. | $\{^{(2S+1)}L_J\}_i$ | Transition | $\{^{(2S+1)}L_J\}_f$ | $P^2$ | $V_0$, MeV | $\gamma$, fm$^{-2}$ | $E_{res}$, keV | $\Gamma_{res}$, keV |
|---|---|---|---|---|---|---|---|---|
| 1 | 2 | 3 | 4 | 5 | 6 | 7 | 8 | 9 |
| 1. | $^2S_{1/2}$ | E1 | $^2P_{3/2}$ | 4 | 0 | 0 | --- | --- |
|    | $^4S_{3/2}$ |    | $^4P_{3/2}$ | 4 | 0 | 0 | --- | --- |
| 2. | $^4D_{1/2}$ | E1 | $^4P_{3/2}$ | 2/5 | 780.0 | 1.0 | --- | --- |
| 3. | $^2D_{3/2}$ | E1 | $^2P_{3/2}$ | 4/5 | 147.705 | 0.25 | 510 | 11 |
|    | $^4D_{3/2}$ |    | $^4P_{3/2}$ | 64/25 |    |    | [510(6)] | [10(10)] |
| 4. | $^2D_{5/2}$ | E1 | $^2P_{3/2}$ | 36/5 | 780.0 | 1.0 | --- | --- |
|    | $^4D_{5/2}$ |    | $^4P_{3/2}$ | 126/25 |    |    |    |    |
| 5. | $^2P_{1/2}$ | M1 | $^2P_{3/2}$ | 4/3 | 780.05 | 1.0 | 146 | 31 |
|    | $^4P_{1/2}$ |    | $^4P_{3/2}$ | 10/3 |    |    | [146(6)] | [---] |
| 6. | $^4P_{5/2}$ | M1 | $^4P_{3/2}$ | 18/5 | 779.03 | 1.0 | 230 | 60 |
|    |    |    |    |    |    |    | [228(10)] | [60(10)] |

In Table 1 the parameters of initial and final states (columns 2 and 4) for various transitions (column 3) and the values of algebraic factors $P^2$ (column 5) in cross sections (2) and (3) have been presented. In columns 6 and 7, the parameters of the initial scattering state potentials and the resonance characteristics obtained with such parameters (columns 8 and 9) are given, and their experimental values are given in square brackets [26].

Since the width for the resonance at 146 keV is not known [26], its potential had been obtained by simply changing the depth from the potential for resonance at 230 keV.

## 6. The total capture cross section and $^{12}$B(*n*,$\gamma$)$^{13}$B reaction rates

On the basis of these results it is possible to calculate nonresonance *E*1 capture No. 1 from Table 1 from the *S* scattering waves with zero depth potential No. 1 without FS to the $^{2+4}P_{3/2}$ GS of $^{13}$B for potentials with FS (10) or (11). The transition occurs from the different *S* scattering waves to the spin-mixed GS of $^{13}$B in the $n^{12}$B channel, so such cross sections are summed [16,17]

$$\sigma(E1) = \sigma(^2S_{1/2} \to\,^2P_{3/2}) + \sigma(^4S_{3/2} \to\,^4P_{3/2}). \tag{12}$$

Here the transitions occur from different partial scattering waves to the same GS of

$^{13}$B, which is matched to the mixed by spin WF $^{2+4}P_{3/2}$ – only coefficients $P_J$ in the expression for the cross section (2) and (3) will be different.

Cross sections of the $E1$ transitions from the $D$ waves with potentials from Table 1 are recorded as averaging [16,17] by spin channel states

$$\sigma(E1) = \{\sigma(^2D_{3/2} \to\, ^2P_{3/2}) + \sigma(^4D_{3/2} \to\, ^4P_{3/2})\}/2 + \\ + \{\sigma(^2D_{5/2} \to\, ^2P_{3/2}) + \sigma(^4D_{5/2} \to\, ^4P_{3/2})\}/2 + \sigma(^4D_{1/2} \to\, ^4P_{3/2}). \quad (13)$$

Cross sections of the $M1$ transitions also are averaged [16,17]

$$\sigma(E1) = \{\sigma(^2P_{1/2} \to\, ^2P_{3/2}) + \sigma(^4P_{1/2} \to\, ^4P_{3/2})\}/2 + \sigma(^4P_{5/2} \to\, ^4P_{3/2}). \quad (14)$$

The results of the total cross section calculations taking into account all transitions from Table 1, are shown in Fig. 1 for the GS potential (6) by the red solid curve. The partial cross sections are shown for illustration the contribution of the resonances on example of the structural red curve:

The black dashed curve in Fig. 1 is the result of calculations of transitions Nos. 1,2,4, i.e., without taking into account resonance states. The green dotted curve shows the results for total cross sections Nos. 5,6 taking into account the resonances at 146 and 230 keV, which corresponds only $M1$ transitions. Violet dashed-dotted curve corresponds to the results for total $E1$ cross sections No. 3 and illuminated the resonance at 510 keV. The analogous calculation of the total summed cross section without splitting to resonance and nonresonance cross sections, but for the GS potential (7), is shown by the blue dashed curve.

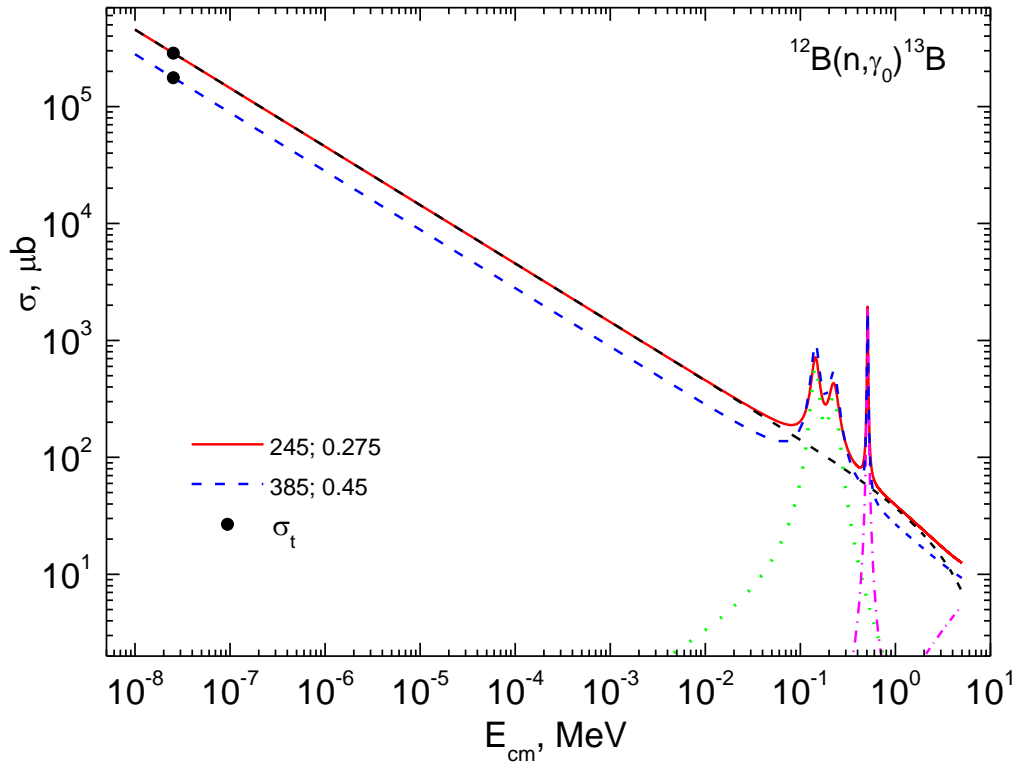

Fig. 1. (Color online) Total cross sections of the neutron radiative capture on $^{12}$B at low energies. The red solid curve shows result for the GS potential (6), blue dashed curve for the GS potential (7). Points are the cross sections calculated at thermal energies for both potentials (6) and (7). All curves are described in text.

In the publications [10,35] the total cross sections of the treating process are not given, so the direct comparison of our results is not possible. In addition, we have not been able to find other results with calculations of such cross sections for any energy range.

If we assume that the GS of $^{13}$B in the $n^{12}$B channel is only the doublet or only the quartet $P$ state, then the total cross sections of the $E1$ transition from the $S$ scattering waves will decrease by half, as follows from the expression (12) and Table 1. Such results were obtained in our preliminary work [27]. In addition, there was only one transition taken into account – from the $S$ scattering wave. This would be possible if the AC of one of these spin states is much smaller than the other. As it is shown above, they have comparable values of 1.79 for the doublet and 1.66 for the quartet states, and the results presented here look the most reasonable.

Another note: we considered the capture only to the GS, but there is information about six excited states (ES), but bound in the $n^{12}$B channel [26]. The possible capture to these ESs can essentially rise up total cross sections comparably with the results shown in Fig. 1. However, we have no data on the AC for these ESs, and therefore there is no possibility to construct for them similar GS interaction potentials and consider processes of such capture.

Since, at the energy range of $10^{-8}$ to $10^{-2}$ MeV, the calculated cross section is almost a straight line (solid red curve in Fig. 1), it can be approximated for the region of these energies by a simple function of the form

$$\sigma_{ap}(\mu b) = \frac{A}{\sqrt{E(\text{MeV})}} \tag{15}$$

with a constant $A = 28.003$ $\mu$b·MeV$^{1/2}$, determining by one point in cross sections at a minimum energy equals $10^{-8}$ MeV. For the solid curve in Fig. 1b, we have $A = 45.412$ $\mu$b·MeV$^{1/2}$. Relative deviation module

$$M(E) = \left|[\sigma_{ap}(E) - \sigma_{theor}(E)]/\sigma_{theor}(E)\right| \tag{16}$$

of calculated theoretical cross section and the approximation of this cross section given by the expression (15) in the range of $10^{-8}$ to $10^{-2}$ MeV is less than 0.5%. From the above equation for approximation, we can obtain an estimate value of the cross section at thermal energy of 25.3 meV (1 meV = $10^{-9}$ MeV) equal to 176.1 mb and 285.5 mb – these results are shown in Fig. 1 by the black dots.

Since there are no experimental data for such reaction, we can try to estimate the cross sections for thermal energy in comparison of the known results for $^{10,11}$B. In [29], new values of thermal cross sections for the neutron capture on $^{10}$B of 394 mb have been reported. As a result of this reaction $^{11}$B is formed, which has the same momentum 3/2$^-$ as $^{13}$B. This fact can be considered as a reason for comparing such cross sections and assume that the thermal cross section for the $^{12}$B($n,\gamma$)$^{13}$B capture can have the same order of magnitude as for the radiative neutron capture on $^{10}$B. Above, we obtained two values of 176 and 285 mb for the thermal cross sections of the reaction under examination, which are quite comparable with 394 mb for n$^{10}$B capture [34].

Furthermore, the reaction rate of the $^{12}$B($n,\gamma$)$^{13}$B capture is considered. It is expressed in cm$^3$mol$^{-1}$sec$^{-1}$ and can be represented in the usual form [20]

$$N_A \langle \sigma v \rangle = 3.7313 \cdot 10^4 \mu^{-1/2} T_9^{-3/2} \int_0^\infty \sigma(E) E \exp(-11.605 E/T_9) dE, \quad (17)$$

where $E$ is specified in MeV; the total cross section $\sigma(E)$ is measured in μb; μ is the reduced mass is given in amu; and $T_9$ is in $10^9$ K. The reaction rate was calculated using 5000 points for total cross sections in the energy range of $10^{-8}$ to 5 MeV with step of 1 keV.

The red solid curve shows the reaction rate depending on the temperature in units of $T_9$ in Fig. 2, which corresponds to the red solid curve in Fig. 1. The blue dashed curve in Fig. 2 shows the reaction rate obtained from the results shown in Fig. 1 by the blue dotted curve. The black dashed curve in Fig. 2 shows the results for the reaction rate for the GS potential (10), excluding resonances. This dashed curve and the red solid curve demonstrate the contribution of the resonances at low energies taken into account in our calculations.

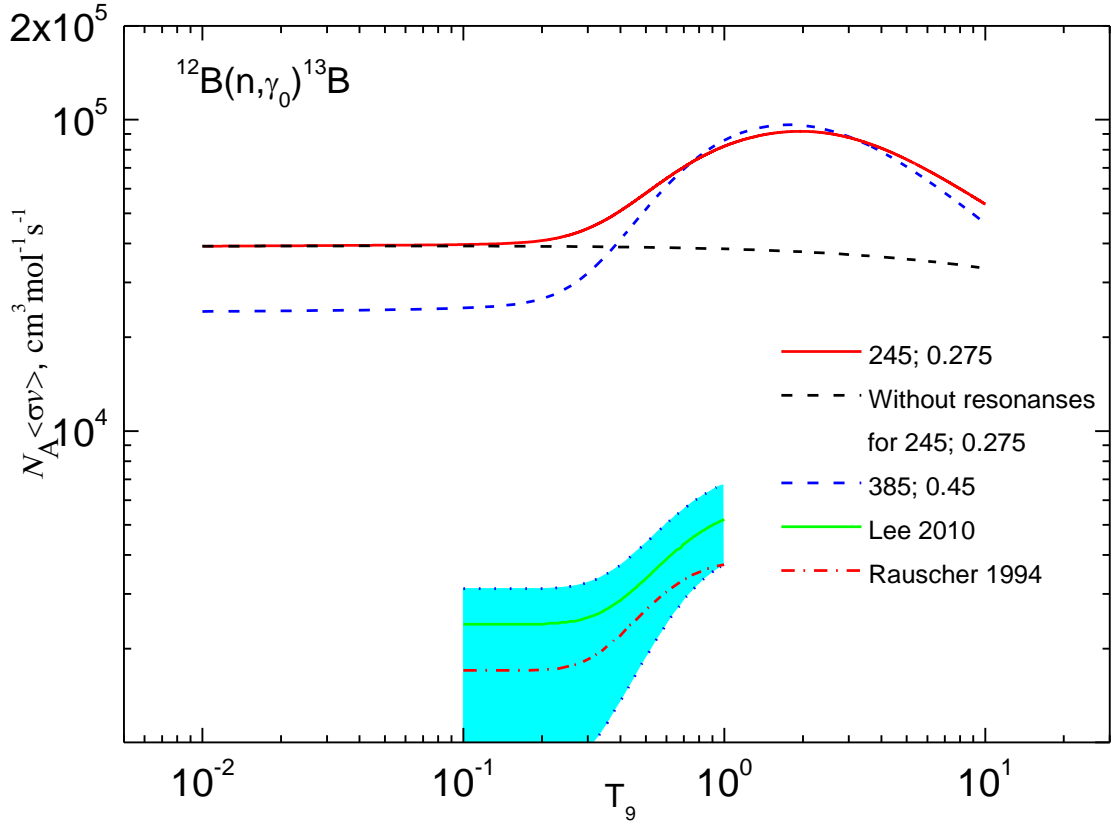

Fig. 2. (Color online) The reaction rate of the neutron radiative capture on $^{12}$B at low energies. The estimation of the reaction rate is taken from [10] and [35]. Curves are calculations with different potentials, the parameters of which are given in the text. Light blue band responds to error band for [10].

At the bottom of Fig. 2, the green dashed curve shows the results obtained in [10], the error range from this work is represented by blue thin dotted curves. The red dashed curve shows the results from the earlier work [35], which is given in [10] to compare the results.

The present result, as can be seen from Fig. 2, for the reaction rate for two different potentials do not fundamentally differ, although at low and medium temperatures the difference reaches to 1.5. However, the difference from the results of [10] and [35] is more than an order of magnitude, although here we took into account only the capture to the GS and to the first three resonance states up to 0.5 MeV.

Then, the calculated reaction rates were parameterized by a function of the form [36]

$$N_A \langle \sigma v \rangle = a_1 / T_9^{b_1} \cdot \exp(a_2 T_9^{b_2}) \cdot (1.0 + a_3 T_9^{b_3} + a_4 T_9^{b_4} + a_5 T_9^{b_5}) + \\ + a_6 / T_9^{b_6} \cdot \exp(-a_7 / T_9^{b_7}) + a_8 / T_9^{b_8} \cdot \exp(-a_9 / T_9^{b_9})$$ (18)

with parameters for the green dotted curve from Table 2 and for the red dotted curve from Table 3. Similar parametrizations were used in [36], but here we also consider the powers $b_i$ at $T_9$ as variational parameters. Such opportunity doubles the number of variable parameters, but allows one to improve the quality of the description of the calculated curve by the analytical approximation (18).

Table 2. Parameters of analytical approximation (18) of the reaction rate of $^{12}$B$(n,\gamma)^{13}$B capture with accounting of the resonances (red solid curve in Fig. 2).

| No. | $a_i$ | $b_i$ |
|---|---|---|
| 1 | 0.1336505E-10 | 0.1211152E+01 |
| 2 | 0.9380520E+00 | 0.2790421E+00 |
| 3 | 0.1535577E+16 | 0.3315706E+00 |
| 4 | 0.2195656E+11 | -0.2160328E-01 |
| 5 | -0.9483378E+15 | 0.1093404E+01 |
| 6 | 0.4260593E+06 | 0.7466873E+00 |
| 7 | 0.1941090E+01 | 0.7387207E+00 |
| 8 | 0.9814834E+05 | 0.1122955E+01 |
| 9 | 0.1336505E-10 | 0.4107266E+00 |
| $\chi^2 = 0.004$ | | |

Table 3. Parameters of analytical approximation (18) of the reaction rate of $^{12}$B$(n,\gamma)^{13}$B without taking into account the resonance transitions (dark dashed curve in Fig. 2).

| No. | $a_i$ | $b_i$ |
|---|---|---|
| 1 | 0.1623266E-11 | 0.1095114E+01 |
| 2 | -0.1476235E+01 | -0.1456108E-02 |
| 3 | 0.6505753E+15 | 0.9294294E+00 |
| 4 | 0.1021481E+16 | 0.9451261E+00 |
| 5 | -0.7519269E+15 | 0.1287142E+01 |
| 6 | 0.4446746E+06 | 0.9552922E+00 |
| 7 | 0.1924944E+01 | 0.8319477E+00 |
| 8 | 0.1028488E+06 | 0.5892998E+00 |
| 9 | 0.1950270E+01 | 0.2592384E+00 |
| $\chi^2 = 0.004$ | | |

The accuracy of parametrization in both cases is of $\chi^2$ equal to 0.004 with 5% errors in the calculated rate.

# 7. The total capture cross section and reaction rates of the $^{12}C(n,\gamma)^{13}C$ process and synthesis of the $^{13}C$ nucleus

In the present time it is admittedly that the stable isotope $^{13}C$ is formed as a result of three presumable processes. These are: $\beta^-$ decay of the nuclide $^{13}B$: the half-life of 17.33(17) ms, emitted energy is 13437.2(11) keV; $\beta^+$ decay of the nuclide $^{13}N$ the half-life of 9.965(4) min and emitted energy is 2220.47(27) keV), and also the $^{12}C(n,\gamma)^{13}C$ capture reaction [17,37]. The processing of the unstable nucleus $^{13}B$ is possible in the processes of the neutron radiative capture reactions on the isotopes $^{10,11,12}B$, consistently. Comparative analysis of the reaction rates for these processes is shown in Fig. 5. The $^{13}N$ nucleus is formed, for example, in the proton capture reaction on $^{12}C$. Besides, synthesis of $^{13}C$ takes place as a result of the direct neutron radiative capture on $^{12}C$. It is reasonable to compare the efficiencies of all these reactions.

Earlier in our works [17,37] we have studied reaction $^{12}C(n,\gamma)^{13}C$ taken into account formation of the isotope $^{13}C$ both in the ground and three low-lying excited states. The partial $(n,\gamma)$ cross sections in comparison with the experimental data are shown in Fig. 3. It is possible in the frame of the MPCM to reproduce all available up to this day data, which is a good argumentation of the adequacy of the reaction rate calculation, carried out here and shown in Fig. 4.

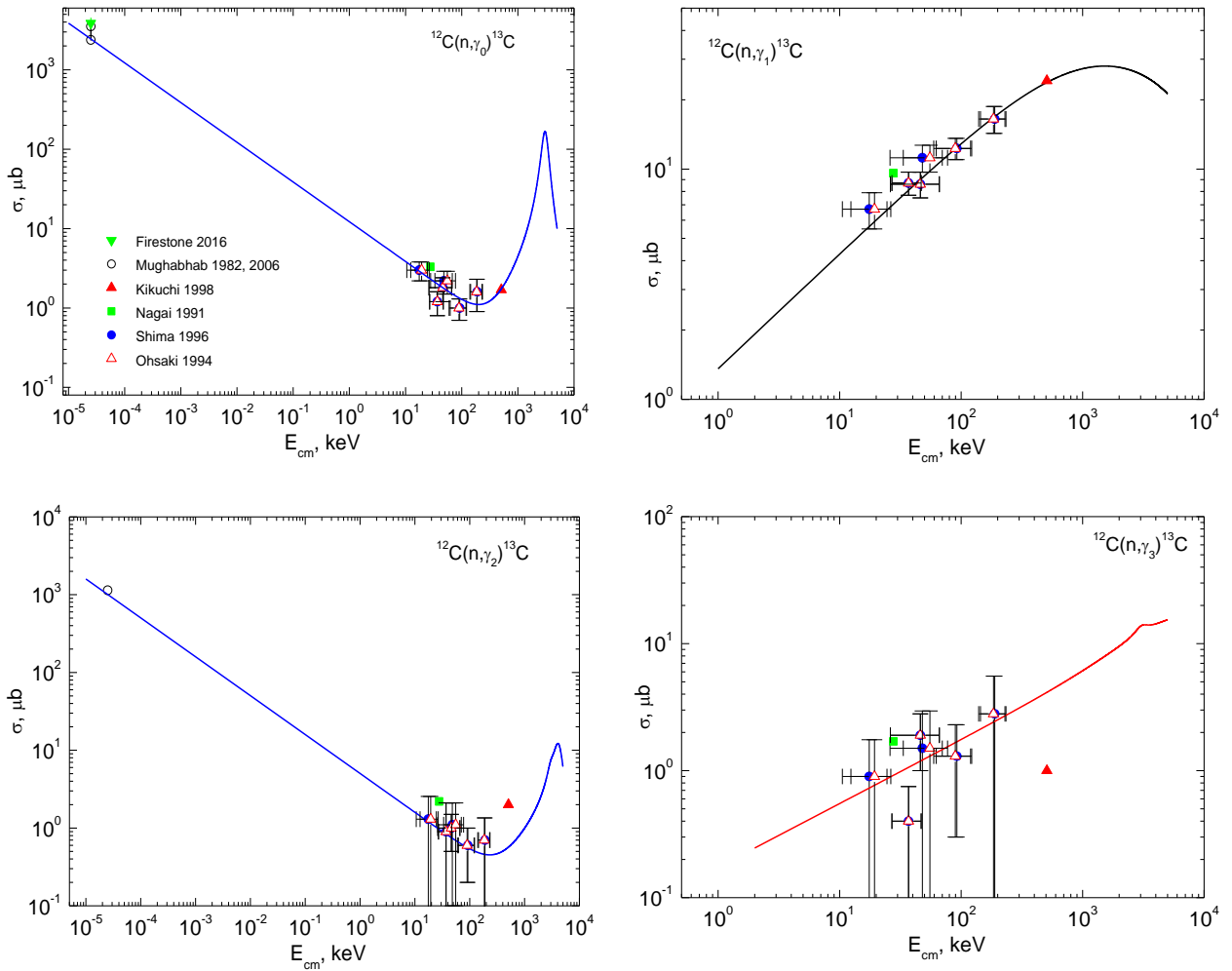

Fig. 3 (Color online) Cross section of the radiative $^{12}C(n,\gamma_{0+1+2+3})^{13}C$ capture. Calculation from [17,37]. Experimental data: ● – [38], ○ – [39], ■ – [40], Δ – [41], ▲ – [42], ▼ – [43].

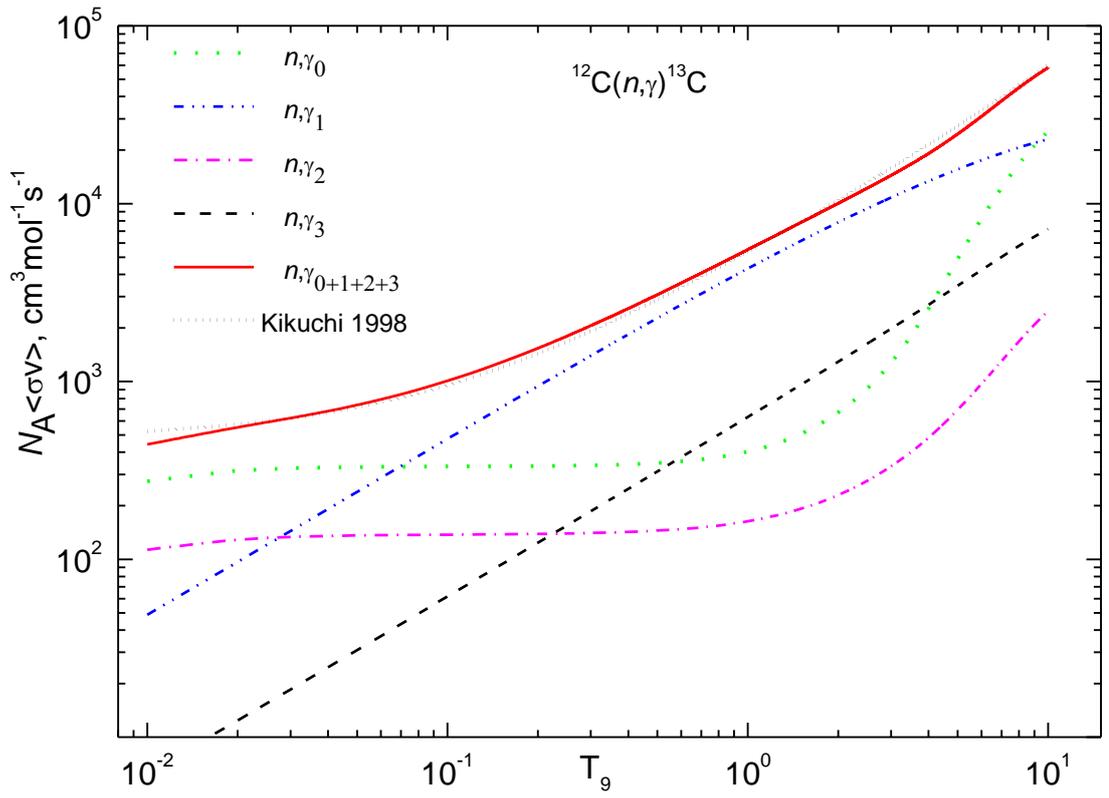

Fig. 4 (Color online) Reaction rate for the $^{12}$C($n,\gamma$)$^{13}$C process. The solid curve is our results; frequent dotted curve is the results [42].

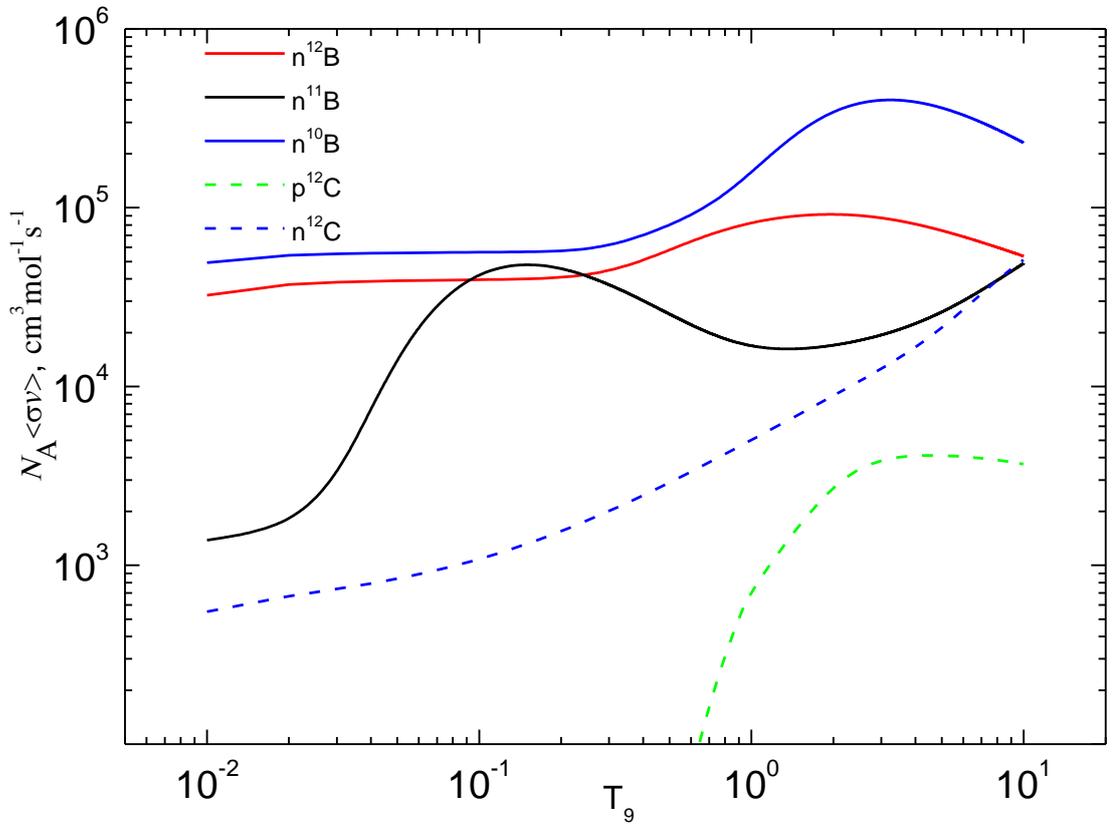

Fig. 5 (Color online) Comparison of reaction rates of the $^{10\text{-}12}$B($n,\gamma$)$^{11\text{-}13}$B, $^{12}$C($n,\gamma_{0+1+2+3}$)$^{13}$C, and $^{12}$C($p,\gamma$)$^{13}$N processes.

The solid curve shows the total capture reaction rate taken into account all excited states. Frequent dotted curve shows the results from [42]. The reaction rate shown in Fig. 4 by the solid curve can be approximated by the function of the form [44]

$$N_A \langle \sigma v \rangle = a_1 / T_9^{2/3} \exp(-a_2 / T_9^{1/3}) \cdot$$
$$(1.0 + a_3 T_9^{1/3} + a_4 T_9^{2/3} + a_5 T_9 + a_6 T_9^{4/3} + a_7 T_9^{5/3} + a_8 T_9^{7/3}) + a_9 / T_9^{1/3}. \quad (19)$$

Table 4. Parameters of approximation (19) of the reaction rate of $^{12}C(n,\gamma_{0+1+2+3})^{13}C$ capture.

| No. | 1 | 2 | 3 | 4 | 5 | 6 | 7 | 8 | 9 |
|---|---|---|---|---|---|---|---|---|---|
| $a_i$ | 19.85505 | 0.5647 | 59.5549 | -0.22945 | -408.1021 | 1259.871 | -514.1163 | 88.10791 | 13.16841 |

Parameters of such approximation are listed in Table 4 and the accuracy of parametrization is $\chi^2 = 0.06$ within 5% error.

The comparison of the reaction rates of the $^{10\text{-}12}B(n,\gamma)^{11\text{-}13}B$, $^{12}C(n,\gamma)^{13}C$, and $^{12}C(p,\gamma)^{13}N$ are shown in Fig. 5 at the temperature range of 0.01 to 10 $T_9$. As it is seen from Fig. 5, the $^{12}B(n,\gamma)^{13}B$ reaction rate prevails over the $^{12}C(n,\gamma_{0+1+2+3})^{13}C$ reaction rate practically by two orders, up to highest temperatures, where they are comparable. At the same time the $^{12}C(p,\gamma)^{13}N$ reaction rate, where the isotope $^{13}N$ is processed, is too small, so that this process cannot be considered as the potential source of stable $^{13}C$.

Results for the neutron capture on $^{12}B$ are same as in Fig. 2 and are shown by the red solid curve. The reaction rate of the neutron capture on $^{11}B$ is taken from our works [45–47] and is shown by the black solid curve. The data on the neutron capture on $^{10}B$ are from our work [15] and shown by the blue solid curve. The $^{12}C(p,\gamma)^{13}N$ reaction rate (green dashed curve) – NACRE II data [48].

## 8. Conclusion

The calculated rates of $^{12}B(n,\gamma)^{13}B$ excess the previous results by approximately to one order. Cross sections and reaction rates of $^{12}C(n,\gamma_{0+1+2+3})^{13}C$ are calculated and compared to the $n^{10}B$, $n^{11}B$, $n^{12}B$, and $p^{12}C$ reaction rates. It is proposed that obtained rates of the $^{12}B(n,\gamma)^{13}B$ reaction should be taken into account in *novel* scenarios of stable isotope $^{13}C$ synthesis without of $^{12}C$ hydrogen burning.

The $^{12}B(n,\gamma)^{13}B$ capture to the GS of $^{13}B$ result to reaction rates that, approximately, higher to an order than the available estimates made in [10] and earlier results from [35]. The $^{12}B(n,\gamma)^{13}B$ capture reaction should be taken into account at the process of prediction ratios for CNO isotopes, in the first place it concerns the $^{12}C/^{13}C$ ratio [11–14], which surely increases in our case.

Our calculations of reaction rates $^{10\text{-}12}B(n,\gamma)^{11\text{-}13}B$ are higher to an order than earlier non-model calculations that allows to state:
1. the $^{12}B(n,\gamma)^{13}B$ reaction is the *dominant* source of synthesis of stable $^{13}C$ isotope via the process of $\beta^-$ decay of $^{13}B$;
2. scenario, based on the evolution of boron isotopes does not lead to the burning of $^{12}C$ contrary to the case of processes $^{12}C(n,\gamma_{0+1+2+3})^{13}C$;

3    the $^{12}$C$(p,\gamma)^{13}$N reaction enters essentially smaller contribution in the balance between hydrogen burning in participation of the $^{12}$C isotope (that is illustrated by Fig. 5) and following synthesis of $^{13}$C in the process $^{13}$N$(\beta^+\nu)^{13}$C.

Thus, the quantitative estimations of $^{13}$C isotope macroscopic accumulation require the additional calculations taking into account the specific conditions for densities of neutrons and protons at different temperature ranges. The obtained results for the $^{12}$B$(n,\gamma)^{13}$B reaction rate show the fundamental effect on the calculation results of the light nuclei efficiencies in different thermonuclear processes of the Universe.

**Acknowledgements**

This work was supported by the Grant of Ministry of Education and Science of the Republic of Kazakhstan through the program BR05236322 "Study reactions of thermonuclear processes in extragalactic and galactic objects and their subsystems" in the frame of theme "Study of thermonuclear processes in stars and primordial nucleosynthesis" through the Fesenkov Astrophysical Institute of the National Center for Space Research and Technology of the Ministry of Digital Development, Innovation and Aerospace Industry of the Republic of Kazakhstan (RK).